\documentclass[%
11pt,
reprint,
onecolumn,
tightenlines,
superscriptaddress,
preprintnumbers,
nofootinbib,
 amsmath,amssymb,amsthm
 physrev,
eqsecnum,tikz,
]{revtex4-2}

\usepackage{isomath}
\usepackage{amsmath,amsthm}
\usepackage{amsbsy}
\usepackage{amssymb}
\usepackage{amscd}
\usepackage{amsfonts}
\usepackage{stmaryrd}
\usepackage{siunitx}
\usepackage{euscript}
\usepackage[utf8]{inputenc}
\usepackage[T1]{fontenc}
\usepackage{newtxtext} 
\everymath{\displaystyle}
\usepackage{exscale}
\usepackage{microtype}
\usepackage{hyperref}
\usepackage{booktabs}
\usepackage{algorithm}
\usepackage{algpseudocode}

\usepackage{graphicx}
\usepackage{boxedminipage}
\usepackage{calc}
\usepackage[usenames,dvipsnames]{xcolor}
\graphicspath{ {media/} }
\usepackage[caption=false,justification=centerlast]{subfig}

\usepackage{setspace}
\usepackage{enumitem}
\setitemize{noitemsep,topsep=0pt,parsep=0pt,partopsep=0pt}
\setenumerate{noitemsep,topsep=0pt,parsep=0pt,partopsep=0pt}
\setdescription{noitemsep,topsep=0pt,parsep=0pt,partopsep=0pt}

\usepackage{soul} % highlighting
\usepackage[normalem]{ulem}

\usepackage{orcidlink}
\usepackage{siunitx}
\usepackage[small]{titlesec}

\titlespacing*{\section}{0pt}{12pt plus 4pt minus 2pt}{2pt plus 2pt minus 2pt}
\titlespacing*{\subsection}{0pt}{12pt plus 4pt minus 2pt}{2pt plus 2pt minus 2pt}
\titlespacing*{\subsubsection}{0pt}{12pt plus 4pt minus 2pt}{2pt plus 2pt minus 2pt}
\titlespacing*{\paragraph}{0pt}{12pt plus 4pt minus 2pt}{2pt plus 2pt minus 2pt}

\makeatletter
    \renewcommand*{\thesection}{\arabic{section}}
    \renewcommand*{\thesubsection}{\thesection.\Alph{subsection}}
    \renewcommand*{\p@subsection}{}
    \renewcommand*{\thesubsubsection}{\thesubsection.\arabic{subsubsection}}
    \renewcommand*{\p@subsubsection}{}
\makeatother

\usepackage{isomath}
\usepackage{amsmath}
\usepackage{amssymb}
\usepackage{amscd}
\usepackage{amsfonts}
\usepackage{upgreek}

\newcommand{\variation}[2]{\updelta_{#2} #1}

\theoremstyle{definition}

\AtEndEnvironment{definition}{\null\hfill\qedsymbol}

\AtEndEnvironment{remark}{\null\hfill\qedsymbol}

\AtEndEnvironment{example}{\null\hfill\qedsymbol}

\AtEndEnvironment{assumption}{\null\hfill\qedsymbol}

\newcommand{\bfzero}{\mathbf{0}}

\DeclareMathOperator{\divergence}{div}

\DeclareMathOperator{\trace}{tr}

\newcommand{\parderiv}[2]{\frac{\partial #1}{\partial #2}}
\newcommand{\dm}{\ \mathrm{d}}
\newcommand{\deriv}[2]{\frac{\dm #1}{\dm #2}}

\newcommand{\bfb}{{\mathbold b}}

\newcommand{\bfg}{{\mathbold g}}

\newcommand{\bfk}{{\mathbold k}}

\newcommand{\bfn}{{\mathbold n}}

\newcommand{\bfq}{{\mathbold q}}

\newcommand{\bfu}{{\mathbold u}}
\newcommand{\bfv}{{\mathbold v}}

\newcommand{\bfx}{{\mathbold x}}

\newcommand{\bfF}{{\mathbold F}}

\newcommand{\bfI}{{\mathbold I}}

\newcommand{\bfK}{{\mathbold K}}

\newcommand{\bfQ}{{\mathbold Q}}

\newcommand{\bfT}{{\mathbold T}}

\everymath{\displaystyle}

\begin{document}

\preprint{To appear in Journal of the Mechanics and Physics of Solids (DOI: \href{https://doi.org/10.1016/j.jmps.2025.106232}{10.1016/j.jmps.2025.106232})}

\title{Impact of Gas/Liquid Phase Change of CO$_2$ during Injection for Sequestration}

\author{Mina Karimi}
    \email{minakari@alumni.cmu.edu}
    \affiliation{Department of Mechanical Engineering, Purdue University}

\author{Elizabeth Cochran}%
    \affiliation{U. S. Geological Survey, Earthquake Science Center}%

\author{Mehrdad Massoudi}%
    \affiliation{National Energy Technology Laboratory}%

\author{Noel Walkington}
    \affiliation{Center for Nonlinear Analysis, Department of Mathematical Sciences, Carnegie Mellon University}

\author{Matteo Pozzi}
    \affiliation{Department of Civil and Environmental Engineering, Carnegie Mellon University}
    
\author{Kaushik Dayal}
    \affiliation{Department of Civil and Environmental Engineering, Carnegie Mellon University}
    \affiliation{Center for Nonlinear Analysis, Department of Mathematical Sciences, Carnegie Mellon University}
    \affiliation{Department of Mechanical Engineering, Carnegie Mellon University}

\date{\today}

%%%%%%%%%%%%%%%%%%%%%
%%%%%%%%%%%%%%%%%%%%%
%%%%%%%%%%%%%%%%%%%%%
%%%%%%%%%%%%%%%%%%%%%

\begin{abstract}
    CO$_2$ sequestration in deep saline formations is an effective and important process to control the rapid rise in CO$_2$ emissions. 
    The process of injecting CO$_2$ requires reliable predictions of the stress in the formation and the fluid pressure distributions -- particularly since monitoring of the CO$_2$ migration is difficult -- to mitigate leakage, prevent induced seismicity, and analyze wellbore stability.
    A key aspect of CO$_2$ is the gas-liquid phase transition at the temperatures and pressures of relevance to leakage and sequestration, which has been recognized as being critical for accurate predictions but has been challenging to model without \textit{ad hoc} empiricisms.
    
    This paper presents a robust multiphase thermodynamics-based poromechanics model to capture the complex phase transition behavior of CO$_2$ and predict the stress and pressure distribution under super- and sub- critical conditions during the injection process.
    A finite element implementation of the model is applied to analyze the behavior of a multiphase porous system with CO$_2$ as it displaces the fluid brine phase.
    We find that if CO$_2$ undergoes a phase transition in the geologic reservoir, the spatial variation of the density is significantly affected, and the migration mobility of CO$_2$ decreases in the reservoir. 
    A key feature of our approach is that we do not \textit{a priori} assume the location of the CO$_2$ gas/liquid interface -- or even if it occurs at all -- but rather, this is a prediction of the model, along with the spatial variation of the phase of CO$_2$ and the change of the saturation profile due to the phase change. 
    
\end{abstract}

\maketitle

%%%%%%%%%%%%%%%%%%%%%
%%%%%%%%%%%%%%%%%%%%%
%%%%%%%%%%%%%%%%%%%%%
%%%%%%%%%%%%%%%%%%%%%

%%%%%%%%%%%%%%%%%%%%%
%%%%%%%%%%%%%%%%%%%%%
%%%%%%%%%%%%%%%%%%%%%
%%%%%%%%%%%%%%%%%%%%%
\section{Introduction} \label{Introduction}

Carbon dioxide (CO$_2$) sequestration and storage in geological formations is among the most promising available approaches to reduce CO$_2$ emissions \cite{NAP25259}. 
The CO$_2$ emitted from different industrial sources can be collected and isolated in deep underground formations ~\cite{chen2017handbook}, or injected into depleted natural gas reservoirs to enhance gas recovery ~\cite{oldenburg2001process}.
Of the various types of geological formations that are considered for CO$_2$ storage, saline aquifers appear to provide the most storage capacity for CO$_2$ ~\cite{ahmad2016injection}. 
However, there remain several fundamental challenges \cite{NAP25259}.

CO$_2$ is typically injected into deep, brine-saturated formations in a dense, supercritical state, as the lower viscosity of supercritical CO$_2$ compared to its liquid phase makes it easier to inject. The higher density of the resident brine drives the injected CO$_2$ to migrate upward toward the top of the reservoir, where it is ideally trapped beneath the caprock. The migration of CO$_2$ and brine involves more complex processes over longer timescales. CO$_2$ can gradually dissolve into the resident brine, forming a denser and more reactive fluid that sinks through convective mixing, reducing the risk of leakage. The CO$_2$-brine interaction is coupled with a series of geochemical reactions with the surrounding subsurface material, leading to alteration in transport properties such as porosity and permeability through dissolution and precipitation processes ~\cite{fu2015rock, ilgen2019coupled, balashov2015reaction, ahmad2016reactive,wang2023geochemically, karimi2024learning}.

%%%%%%%%%%%%%%%%%%%%%
%%%%%%%%%%%%%%%%%%%%%
%%%%%%%%%%%%%%%%%%%%%
%%%%%%%%%%%%%%%%%%%%%
\paragraph*{Prior Work.} 

Reliable predictions, based on numerical simulations, are needed to predict CO$_2$ propagation in deep underground formations and estimate the capacity of storage to reduce possible leakage risks \cite{vilarrasa2019induced}. 
Further, injecting supercritical CO$_2$ into deep geological formations can increase fluid pressure near injection wells, alter the stress regime along pre-existing faults, potentially leading to fault activation and induced seismicity \cite{rutqvist2016fault, vilarrasa2019induced}. 
Such seismic events can compromise the integrity of the underground reservoir and caprock, posing a risk of CO$_2$ leakage to shallower layers \cite{chang2019coupled, newell2020numerical}. Consequently, accurately predicting the spatial distribution of pressure, CO$_2$ phase, and density is essential for ensuring the safety of CO$_2$ storage.

Several studies have proposed numerical models to investigate the subsurface transport of injected CO$_2$, focusing primarily on fluid transport while often neglecting the effect of deformation of the porous medium ~\cite{hosseini2012analytical, huang2015parallel, giorgis20072d, preisig2011coupled, sun2018flow}. 
These models usually simplify fluid saturation-pressure relationships and CO$_2$ properties, commonly assuming CO$_2$ to remain in a supercritical state at constant temperatures ~\cite{negara2011simulation, rabinovich2015upscaling, song2014analytical, nordbotten2010analysis}. 
While other studies account for coupled fluid transport and matrix deformation, they still depend on simplified constitutive models for pressure-saturation relations and fluid properties ~\cite{huang2015parallel, preisig2011coupled, okwen2011temporal, vilarrasa2016two, piao2018dynamic, kim2018co2, kim2021numerical, singh2011non}

In general, simulation of phase transformation in fluids is challenging because of the significant change in the density and viscosity values, the evolving interfaces between liquid and gas, and the complex behavior of the liquid/gas mixture. 
Recently, free energy-based approaches have been proposed to predict the phase transformation of fluids for different applications. 
For instance, Ateshian and Shim developed an energy-based formulation with particular emphasis on the jump conditions across the interface of liquid and gas phases \cite{ateshian2022continuum}; Hu et al. proposed a formulation based on the Navier-Stokes-Korteweg equations to simulate the liquid-vapor phase transition under non-equilibrium conditions \cite{hu2022direct}; and we presented a variational energy-based model for multi-phase flow which models the behavior of phase-changing fluids in a porous medium \cite{karimi2022energetic}.

Most prior studies of CO$_2$ sequestration ignore the CO$_2$ phase transition to simplify the flow analysis; however, it can cause considerable error, specifically for short-term, high-rate injections and in shallower reservoirs. 
Some studies consider the phase transition of CO$_2$ during injection and can be broadly categorized into three approaches.
The first approach assumes a constant density for the liquid phase of CO$_2$ and typically combines the empirical Brooks-Corey model, that provides a relationship between fluid pressure and phase saturation, with the equation of state for CO$_2$ e.g., \cite{kolditz2012numerical, goerke2011numerical}; this assumption can introduce considerable errors in predicting pressure and density profiles.
The second approach employs a multi-phase flow by considering phase partitioning criteria to simulate the CO$_2$ phase change. This approach considers a pressure threshold to define different phases (e.g., TOUGH simulators \cite{pruess2004numerical,pruess2011integrated,pruess2011eco2m}); however, the use of a sharp threshold in the subcritical regions, where CO$_2$ pressure is close to the critical pressure, can lead to unstable behavior, abrupt changes in properties, and CO$_2$ phase change with small perturbations. 
The third approach focuses on studying the phase change of CO$_2$ during injection by using multi-phase flow models to simulate the behavior of CO$_2$ in both its liquid and gas states and accounting for the phase transition of CO$_2$ by considering the variation in the enthalpy of the CO$_2$ liquid and gas phases, which requires the calculation of internal energy for each phase, e.g. \cite{lu2014transient,wan2021modeling,sasaki2011heat}; however, these models are typically limited to simulating the behavior of CO$_2$ in the injector.

%%%%%%%%%%%%%%%%%%%%%
%%%%%%%%%%%%%%%%%%%%%
%%%%%%%%%%%%%%%%%%%%%
%%%%%%%%%%%%%%%%%%%%%
\paragraph*{Contributions of This Paper.}

Typically, CO$_2$ is stored in the supercritical phase in deep saline formations at depths of \SI{800}{m} to \SI{3}{km}.  
However, in a large reservoir, CO$_2$ can migrate to shallower depths ~\cite{van2009fluid}, and also leakage of CO$_2$ from the reservoir to faults or abandoned wellbores and upward migration of CO$_2$ to the ground surface can decrease the temperature and pressure and lead to subcritical conditions ~\cite{pruess2011integrated, pruess2004numerical, vilarrasa2017thermal}. 
Sub-critical conditions of CO$_2$ at depths shallower than $500-750$ \SI{}{m} can provide a mixture of CO$_2$ phases (gas, liquid, and super-critical), creating a complicated multi-phase flow process.

In this work, we investigate the complex behavior of multiphase CO$_2$ flow that consistently accounts for the CO$_2$ phase transition using an approach based on the thermodynamical free energy \cite{karimi2022energetic}.
Our approach enables accounting of the entire mixture of CO$_2$ phases without any \textit{ad hoc} assumptions on the behavior of CO$_2$ and using a minimal set of state variables.
We use the simple yet effective van der Waals (vdW) model to simulate the complex behavior and phase transition of CO$_2$ under sub-critical conditions. 
The free-energy-based formulation and the vdW model allow us to simulate the transition from the gas to the liquid phase of CO$_2$ at different temperatures. This approach consistently provides fluid pressure-density relations without relying on additional \textit{ad hoc} assumptions, therefore preventing unstable behavior and oscillations during the phase transition.
We study the impact of the CO$_2$ phase transition on the pressure, saturation, density distributions, and migration mobility of CO$_2$ in a geological formation. We also investigate the upward mobility of gas-liquid CO$_2$ and supercritical CO$_2$ in the event of leakage.
In Appendix \ref{sec:comparison}, we briefly compare the approach proposed in this paper with conventional multiphase methods.

%%%%%%%%%%%%%%%%%%%%%
%%%%%%%%%%%%%%%%%%%%%
%%%%%%%%%%%%%%%%%%%%%
\section{Model Formulation}\label{sec:co2 model formulation}

We consider CO2 injected into a saturated saline reservoir containing incompressible brine with a deformable solid skeleton. To simulate this multiphase fluid unsaturated system, we use the energetic formulation developed by Karimi et al. ~\cite{karimi2022energetic}. 
We assume that our system consists of three distinct immiscible components: the solid skeleton, CO$_2$, and the brine fluid phases, indexed by subscripts $s$, $c$, and $b$, respectively. 
For simplicity, we neglect chemical reactions and assume isothermal conditions. 

The overall structure of our approach is variational; while different from the usual Coleman-Noll procedure, it provides an approach that is equally consistent with thermodynamics \cite{naghibzadeh2025accretion}.
We begin with a free energy that is formulated in Section \ref{sec:Energetics} and has contributions from the solid skeleton as well as the fluid phases.
From this energy, the mechanical response, corresponding to momentum balance, is obtained by setting to zero the variation of the energy with respect to the deformation.
The fluid response is obtained by defining the chemical potential as the variation of the energy with respect to the fluid density, and then relating to fluid velocity to the gradient of the chemical potential.
Finally, the variation of the energy with respect to the volume fraction provides the local balance of fluid pressure, i.e., at a given spatial location, the pressure must be equal in all of the fluid phases.
Finally, we discuss briefly in Section \ref{sec:dissipation} the non-negativity of the dissipation.

%%%%%%%%%%%%%%%%%%%%%
%%%%%%%%%%%%%%%%%%%%%
\subsection{Kinematics and Notation} \label{sec:Kinematics}

We use the subscript $\cdot_0$ to refer to quantities and differential operators in the reference configuration.  
We define the deformation through the referential position $\bfx_0$, the deformed position $\bfx(\bfx_0,t)$, and the displacement $\bfu(\bfx_0,t):=\bfx(\bfx_0,t)-\bfx_0$.
Then, we have the deformation gradient tensor $\bfF=\parderiv{\bfx}{\bfx_0}$ and the Jacobian $J = \det \bfF$.

We define two densities for each fluid phase $i$: the mass of the fluid phase per unit deformed volume of the entire mixture, denoted $\mathcal{R}_i$; and the mass of the fluid phase per unit deformed volume occupied by that phase, denoted the ``true'' density $\rho_{i}$.
These are related by $\mathcal{R}_i = \phi_i \rho_i$, where $\phi_i$ is the volume fraction of phase $i$. 
The relations between referential and current quantities, following the assumption of affine transformation ~\cite{karimi2022energetic, gajo2010general, li2004dynamics, hong2008theory}, are $\mathcal{R}_{i} = J^{-1} \mathcal{R}_{0i}$ and $\rho_i = \frac{\phi_{0i}}{\phi_{i}} J^{-1} \rho_{0i}$.

%%%%%%%%%%%%%%%%%%%%%%%%%%%%%%%%%%%%%%
%%%%%%%%%%%%%%%%%%%%%%%%%%%%%%%%%%%%%%

\subsection{Energetics} \label{sec:Energetics}

The thermodynamic free energy of the system is defined as:
\begin{equation}
\label{eqn:energy}
    \mathcal{E}\left[\bfx, \mathcal{R}_{0c}, {\phi}_{c}\right] 
    = 
    \int_{\Omega_0} \left( 
        W_{0s} \left( \bfF \right) + W_{0c} \left( \mathcal{R}_{0c},{\phi}_c, J \right) + \frac{\epsilon}{2}|\nabla_0 \mathcal{R}_{0c}|^2 - \bfb_0\cdot\bfx
    \right) \dm \Omega_0
\end{equation}
where $W_{0s}$ and $W_{0c}$ are the (Helmholtz) free energies per unit referential volume of the skeleton and the CO$_2$, respectively; the brine is assumed to be incompressible, and therefore, its free energy does not appear.
The potential due to gravity is given by $\bfb_0\cdot\bfx$, where $\bfb_0 = \mathcal{R}_0\bfg = \left( \mathcal{R}_{0s} +\mathcal{R}_{0c} + \mathcal{R}_{0b} \right) \bfg$ is the body force due to gravity $\bfg$.
The gradient term with the coefficient $\epsilon$ is a phase-field-like regularization to account for the surface energy of the CO$_2$ gas/liquid interface and provide smoothing to allow the use of straightforward computational methods that do not need to track free boundaries \cite{fu2016thermodynamic}.
This energy includes contributions due to the deformation of the solid skeleton as well as due to changes in volume of the (compressible) fluid phases.

The solid skeleton is modeled using a simple compressible neo-Hookean free energy:
\begin{equation} \label{eq:Neo-Hookean}
    W_{0s}(\bfF)= \phi_{s} \left( \frac{\mu}{2} (\trace(\bfF^T \bfF)-2) - \mu \log{J}+ \frac{\lambda}{2}(\log{J})^2 \right)
\end{equation}
where $\mu$ and $\lambda$ are the Lame elastic constants. 
The CO$_2$ is modeled using the vdW free energy in the setting of hyperelasticity by the expression \cite{karimi2022energetic,abeyaratne2006evolution}:
\begin{equation}\label{eqn:Van der Waal}
    W_{0c}(\mathcal{R}_{0c}, {\phi}_c, J)=c \mathcal{R}_{0c} \bar{R}T\left( 1-\log(c\bar{R}T)\right)- \mathcal{R}_{0c} \bar{R}T\log\left( \frac{J{\phi}_c}{\mathcal{R}_{0c}}-b\right) -a\frac{\mathcal{R}_{0c}^2}{J{\phi}_c}
\end{equation}
where $\bar{R}$ is the ideal gas constant; $T$ is the temperature; $c$ is a non-dimensional constant, and $a$ and $b$ are constants that relate to the phase transition.

While we have made specific constitutive choices above, the proposed framework is general and allows for the use of different free energy models. 
In Appendix \ref{sec:peng}, we compare the simple vdW model against the for an explanation of simulating CO$_2$ phase using the Peng-Robinson free energy model.

%%%%%%%%%%%%%%%%%%%%%
%%%%%%%%%%%%%%%%%%%%%
\subsection{Incompressibility of Brine}
\label{sec:incomp-brine}

The incompressibility of brine introduces a geometric constraint.
The volume fractions of CO$_2$, skeleton, and brine must satisfy ${\phi}_s + {\phi}_c + {\phi}_b = 1$.
Since the solid phase deforms affinely, we have that ${\phi}_s = {\phi}_{0s}$, i.e., the volume fraction in the deformed configuration is equal to the volume fraction in the reference configuration.
For the incompressible brine, we require that the true density in the current configuration have a fixed value ${\rho}_b$, giving ${\phi}_b = \frac{\mathcal{R}_{0b}}{J{\rho}_b}$.
Together, these provide the relation ${\phi}_c + {\phi}_{0s} + \frac{\mathcal{R}_{0b}}{J {\rho}_b} = 1 $.
We enforce the constraint by using a Lagrange multiplier $p$ that corresponds to the fluid pressure in the brine.
We define the Lagrangian functional as follows:
\begin{equation} \label{eq:Lagrangian}
    \mathcal{L} [ \bfx, \mathcal{R}_{0c}, \mathcal{R}_{0b}, \Tilde{\phi}_c, p ] = \int_{\Omega_0} \left( W_{0s}\left( \bfF \right) + W_{0c}\left( \mathcal{R}_{0c}, \Tilde{\phi}_c \right) + \frac{\epsilon}{2}|\nabla_0 \mathcal{R}_{0c}|^2 - \bfb_0\cdot\bfx + p\left( \Tilde{\phi}_c +J \phi_{0s} + \frac{\mathcal{R}_{0b}}{\rho_b} - J \right) \right) \dm \Omega_0
\end{equation}
For mathematical simplicity, we define $\Tilde{\phi}_c = J\phi_c$ as a primary variable.

%%%%%%%%%%%%%%%%%%%%%
%%%%%%%%%%%%%%%%%%%%%
\subsection{Fluid Transport} \label{sec:Balance of mass}

We obtain the chemical potentials of the CO$_2$ and brine by taking the variational/functional derivative of the Lagrangian (\ref{eq:Lagrangian}) with respect to $\mathcal{R}_{0i}$ ~\cite{gajo2010general, coussy2004poromechanics, karimi2022energetic}:
\begin{equation}
    \label{eq:eta co2}
    \eta_{0c} = -\parderiv{{W}_{0c}}{\mathcal{R}_{0c}}+\bfg\cdot\bfx - \epsilon\divergence_0(\nabla_0 \mathcal{R}_{0c}) 
    \qquad \text{ and } \qquad
    \eta_{0b} = -\frac{p}{\rho_b}+\bfg\cdot\bfx
\end{equation}
The referential relative velocity vector for each fluid phase is defined as $\bfv_{0i} = \bfK_i \nabla_0\eta_{0i}$, where $\bfK_i$ is the referential permeability.
The permeability in the current configuration is $\bfk_i = J^{-1}\bfF\bfK_i\bfF^\top$, and $\bfk_i = \frac{\kappa}{\gamma_i}\rho_i\bfI$ with $\kappa$ the true permeability; $\gamma_i$ the dynamic viscosity of the fluid; and $\bfI$ is the second-order identity tensor. The viscosity of each fluid phase is assumed to be constant. 
Consequently, using the relation $\bfq_{0i} = \mathcal{R}_i \bfv_{0i}$ for the referential fluid flux vector, we have the flux vectors for the CO$_2$ and brine:
\begin{align}\label{eq:flux}
    \bfq_{0c} = -\bfK_c \left( \mathcal{R}_c\nabla_0\parderiv{{W}_{0c}}{\mathcal{R}_{0c}}- \mathcal{R}_c\bfF^\top\bfg + \epsilon \mathcal{R}_c \nabla_0\divergence_0(\nabla_0 \mathcal{R}_{0c})\right) 
    \qquad \text{ and } \qquad
    \bfq_{0b} = -\bfK_b \left( \phi_b\nabla_0 p - \mathcal{R}_b\bfF^\top\bfg \right)
\end{align}

Finally, using the conservation of mass for each fluid phase, we can write the governing PDE:
\begin{equation}
\label{eqn:transport}
    -\int_{\partial\Omega} \bfq_i \cdot\bfn \dm S = \deriv{\ }{t} \left(\int_{\Omega} \mathcal{R}_i \dm V\right)
    \implies
    - \divergence_0 \bfq_{0i} = \deriv{\ }{t} \mathcal{R}_{0i} 
\end{equation}
which is the generalization of the standard Darcy law, which is the simplest model for fluid transport in rigid porous media.

We highlight that the surface energy contribution leads to a third-order derivative in the CO$_2$ flux vector in \eqref{eq:flux}$_1$ and, consequently, a fourth-order derivative in the balance of mass for the CO$_2$ phase \eqref{eqn:transport}. 
To be able to use a standard finite element approach, we use a mixed method by introducing $\psi := \divergence_0 \left( \nabla_0 \mathcal{R}_{0c}\right)$, as discussed further in Section \ref{sec:numerics}.

%%%%%%%%%%%%%%%%%%%%%
%%%%%%%%%%%%%%%%%%%%%
\subsection{Balance of Linear Momentum} \label{Balance of momentum}

Setting to zero the variational derivative of the Lagrangian functional \eqref{eq:Lagrangian} with respect to $\bfx$ gives the balance of momentum:
\begin{equation}
\label{eqn:mom-balance}
    \divergence_0 \bfT  + \bfb_0 = \bfzero
\end{equation}
where we have defined $\bfT:=\frac{\partial W_{0s}}{\partial \bfF} + \frac{\partial W_{0c}}{\partial \bfF}$, the first Piola stress tensor.
This corresponds to the total stress and consists of contributions from the elasticity of the solid skeleton and from the volume changes of the compressible fluids.
For the constitutive choices in \eqref{eq:Neo-Hookean} and \eqref{eqn:Van der Waal}, we have the expression:
\begin{equation}
    \bfT = \phi_{0s} \left( \mu\bfF -\mu\bfF^{-\top} +\lambda \log{J} \bfF^{-\top}\right)-(1-\phi_{0s})pJ\bfF^{-\top}
\end{equation}

The balance of angular momentum is automatically satisfied by the choice of a frame indifferent energy density in \eqref{eqn:energy}.
That is, since the energy densities in \eqref{eq:Neo-Hookean} and \eqref{eqn:Van der Waal} are invariant under the transformation $\bfF \to \bfQ \bfF$, for every rotation $\bfQ$, it follows that the stress $\bfT:=\frac{\partial W_{0s}}{\partial \bfF} + \frac{\partial W_{0c}}{\partial \bfF}$ satisfies the symmetry requirement $\bfT \bfF^{\top} = \bfF \bfT^{\top}$.

%%%%%%%%%%%%%%%%%%%%%
%%%%%%%%%%%%%%%%%%%%%
%%%%%%%%%%%%%%%%%%%%%
\subsection{Balance of Fluid Pressure}

To find the balance of fluid pressure, we set the variation of the Lagrangian (\ref{eq:Lagrangian}) with respect to $\tilde{\phi}_c$ to $0$, while imposing the constraint that the volume fractions must sum to $1$ (Section \ref{sec:incomp-brine}).
This provides the relation:
\begin{equation}
    \phi_{0c} \parderiv{W_{0c}}{\phi_c} = \phi_{0b} \parderiv{W_{0b}}{\phi_b}
\end{equation}

Using the relation that the pressure is related to the Helmholtz free energy density per unit volume by $p = W(\rho) -\rho \deriv{W}{\rho}$, we have that $\phi_{0c} \parderiv{W_{0c}}{\phi_c} = J p_{0c}$ and $\phi_{0c} \parderiv{W_{0b}}{\phi_b} = p + p_{0b}$ \cite{chua_deformation_2024}, where $p_{0b}$ is the initial pressure of brine.
Combining these relations results in the balance of fluid pressure, i.e., the fluid pressure in the brine is equal to the fluid pressure in the CO$_2$.
For the vdW model, we can write this explicitly.
\begin{equation}
\label{eqn:fluid-balance}
    -\frac{RT\rho_c}{1-b\rho_c} + a\rho_c^2 +p + p_{0b}= 0
\end{equation}

We note that we have ignored capillary pressure effects but including these effects would not change the overall structure of our argument.

%%%%%%%%%%%%%%%%%%%%%
%%%%%%%%%%%%%%%%%%%%%
\subsection{Non-negativity of Dissipation}
\label{sec:dissipation}

To ensure compatibility with thermodynamics, it is sufficient in the isothermal setting to ensure that the dissipation is non-negative for any process \cite{penrose1990thermodynamically,abeyaratne2006evolution}.
Following \cite{de2018disclinations,agrawal2015dynamic,agrawal2015dynamic-2,naghibzadeh2025accretion}, we compute the time derivative of the energy from \eqref{eqn:energy}:
\begin{equation}
    \deriv{\mathcal{E}}{t} = \int_{\Omega_0} \left( \variation{\mathcal{E}}{\bfx} \deriv{\bfx}{t} + \variation{\mathcal{E}}{\mathcal{R}_{0c}} \deriv{\mathcal{R}_{0c}}{t} + \variation{\mathcal{E}}{{\phi}_{c}} \deriv{{\phi}_{c}}{t}\right) \dm\Omega_0    
\end{equation}
where $\variation{\mathcal{E}}{(\cdot)}$ is the variational derivative of $\mathcal{E}$ with respect to $(\cdot)$.

In our approach, we have set $\variation{\mathcal{E}}{\bfx} \equiv \bfzero$ to obtain linear momentum balance, and $\variation{\mathcal{E}}{{\phi}_{c}} \equiv 0$ t obtain fluid pressure balance.
Further, substituting for $\deriv{\mathcal{R}_{0c}}{t}$ from \eqref{eqn:transport} and using that $\bfq_{0c} = \mathcal{R}_{c} \bfv_{0c} = - \bfK_c \nabla_0 \variation{\mathcal{E}}{\mathcal{R}_{0c}}$, we can write:
\begin{equation}
    \deriv{\mathcal{E}}{t} 
    = 
    \int_{\Omega_0} \variation{\mathcal{E}}{\mathcal{R}_{0c}} \deriv{\mathcal{R}_{0c}}{t} \dm\Omega_0    
    =
    \int_{\Omega_0} \variation{\mathcal{E}}{\mathcal{R}_{0c}} \divergence_0 \left( \bfK_c \nabla_0 \variation{\mathcal{E}}{\mathcal{R}_{0c}} \right) \dm\Omega_0
    =
    - \int_{\Omega_0} \nabla_0 \variation{\mathcal{E}}{\mathcal{R}_{0c}} \cdot \bfK_c \nabla_0 \variation{\mathcal{E}}{\mathcal{R}_{0c}} \dm\Omega_0
\end{equation}
where we have used integration-by-parts to obtain the final expression.
From the positive-definiteness of $\bfK_c$ we have that the integrand is non-negative pointwise above, leading to the conclusion that $\deriv{\mathcal{E}}{t} \leq 0$ for every process, in accord with thermodynamics.

%%%%%%%%%%%%%%%%%%%%%
%%%%%%%%%%%%%%%%%%%%%
%%%%%%%%%%%%%%%%%%%%%

\subsection{Weak Form and Finite Element Implementation}
\label{sec:numerics}

Our numerical solution is performed using the Finite Element Method in the open-source framework FEniCS \cite{fenicsprojectFEniCS}.

We have six unknowns: $\bfx$, $\mathcal{R}_{0c}$, $\mathcal{R}_{0b}$, $\tilde{\phi}_c$, $p$, and $\psi$. 
We denote the corresponding test functions by $\hat{\bfu}$, $\hat{\mathcal{R}}_c$, $\hat{\mathcal{R}}_b$, $\hat{\Phi}$, $\hat{p}$, and $\hat{\Psi}$. 
These are governed by 3 PDEs, 2 pointwise constraints, and one substitution to provide a mixed method to deal with higher-order derivatives.
The weak forms are as follows:
\begin{align}
    \text{Momentum Balance \eqref{eqn:mom-balance}: } 
    & 
    \int_{\Omega_0} \left( -\bfT \cdot \nabla_0 \hat{\bfu} + \bfb_0\cdot \hat{\bfu} \right) \dm \Omega_0 + \int_{\partial \Omega_0} \left( \bfT \bfn \right)\cdot \hat{\bfu} \dm S = \bfzero
    \\
    \text{CO2 Transport \eqref{eqn:transport}: }    
    & \int_{\Omega_0} \left( -\bfq_{0c}\cdot\nabla_0 \hat{\mathcal{R}}_c + \frac{\mathcal{R}_{0c}^n - \mathcal{R}_{0c}^{n-1}}{\dm t} \hat{\mathcal{R}}_c \right) \dm \Omega_0 + + \int_{\partial \Omega_0} \left( \bfq_{0c}\cdot \bfn \right) \hat{\mathcal{R}_c} \dm S = 0
    \\
    \text{Brine Transport \eqref{eqn:transport}: }    
    & \int_{\Omega_0} \left( -\bfq_{0b}\cdot\nabla_0 \hat{\mathcal{R}}_b + \frac{\mathcal{R}_{0b}^n - \mathcal{R}_{0b}^{n-1}}{\dm t} \hat{\mathcal{R}}_b \right) \dm \Omega_0 + + \int_{\partial \Omega_0} \left( \bfq_{0b}\cdot \bfn \right) \hat{\mathcal{R}_b} \dm S = 0
    \\
    \text{Incompressibility of Brine \eqref{eq:Lagrangian}: }    
    & \int_{\Omega_0} \left( J \phi_{0s} +\tilde{\phi}_c + \frac{\mathcal{R}_{0b}}{\rho_b} -J \right) \hat{\Phi} \dm \Omega_0 = 0 
    \\
    \text{Fluid Pressure Balance \eqref{eqn:fluid-balance}: }    
    & \int_{\Omega_0} \left( -\frac{RT\rho_c}{1-b\rho_c} + a\rho_c^2 +p + p_{0b} \right) \hat{p} \dm \Omega_0 = 0
    \\
    \text{The substitution $\psi := \divergence_0 \left( \nabla_0 \mathcal{R}_{0c}\right)$: }
    & \int_{\Omega_0} \left( \psi \hat{\Psi} + \nabla_0 \mathcal{R}_{0c} \cdot \nabla_0 \hat{\Psi} \right) \dm \Omega_0 - \int_{\partial \Omega_0} \hat{\Psi} \nabla_0 \mathcal{R}_{0c}\cdot\bfn \dm S = 0
\end{align}
where $\bfn$ is the unit normal to the boundary $\partial \Omega_0$.

We use a triangular mesh with roughly constant refinement throughout the domain.
We use continuous interpolations of order $1$ for $\bfx$, $\hat{\mathcal{R}}_{0c}$, $\hat{\mathcal{R}}_{0b}$, and $\tilde{\phi}_c$; discontinuous Galerkin (DG) interpolations of order $0$, which are piecewise constant within each element, for $p$; and DG interpolations of order $1$, which are piecewise linear, for $\psi$.
This follows the general heuristic that Lagrange multipliers are interpolated at lower-order to satisfy the inf-sup condition \cite{Strang-CSE}.

For time evolution, we use an implicit Euler finite-difference scheme and use Newton methods to solve the resulting nonlinear problem at each time step.

%%%%%%%%%%%%%%%%%%%%%
%%%%%%%%%%%%%%%%%%%%%
%%%%%%%%%%%%%%%%%%%%%

\section{Numerical Results}\label{sec:co2 numerical analysis}

\subsection{CO$_2$ Injection into an Underground Layer}

We consider a model situation of CO$_2$ injection from a well into an underground saline layer and find that CO$_2$ propagation into the layer slows down if phase transformations occur.

\begin{figure}[htb!]
    	\includegraphics[width=0.55\textwidth]{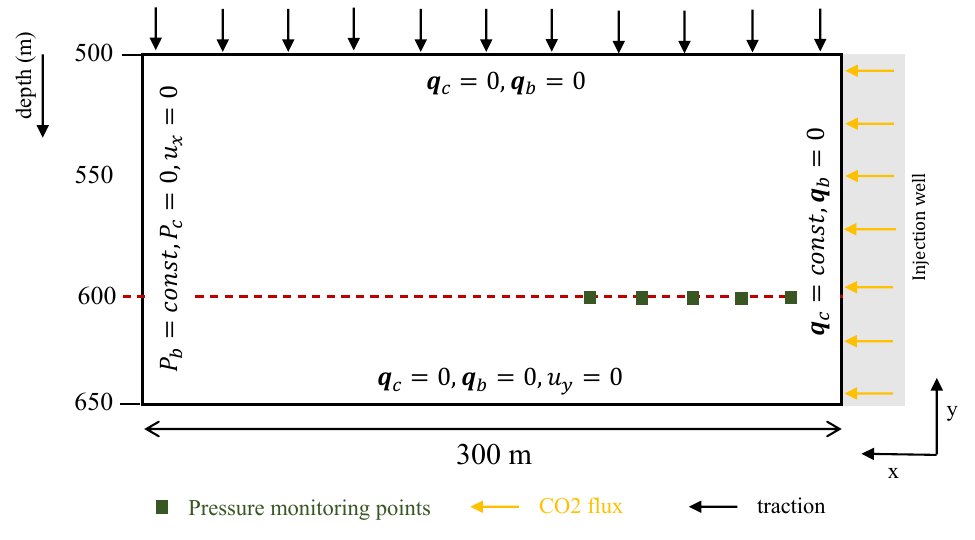}
        \caption{The geometry of the porous layer and injection well. }
    	\label{fig:fig2}
\end{figure}

We consider a porous layer at the depth of $500$ \SI{}{m}, which is initially saturated with brine at an initial pressure of $p_0 = $ \SI{6}{MPa}. We consider porosity of $0.2$ ($\phi_s = 0.8$), and inject super-critical CO$_2$ from the right boundary with constant rate $q_c = 0.088$ \SI{}{kg/m^2.s}. We assume a symmetric domain in the left and right side of the injection well with the geometry and boundary conditions shown in Figure \ref{fig:fig2}, and the material properties of the solid skeleton, brine, and CO$_2$ are listed in Table \ref{tab:properties}. We use $\epsilon = 0.01$.

\begin{table*}[h!]
    \centering
    \begin{tabular}{l l}%{|| c | c r c c c||} 
        \hline
        Property & Value\\
        \hline
        Solid phase Lame constant, $\lambda$ \qquad \qquad & \SI{144.2}{\mega\pascal}\\
        Solid phase Lame constant, $\mu$ & \SI{96.1}{\mega\pascal}\\
        Solid density, $\rho_s$ & \SI{2000}{kg/m^3}\\
        Intrinsic permeability, $\kappa$ & \SI{2e-12}{m^2}\\
        Brine density, $\rho_b$ & \SI{1100}{kg/m^3}\\
        Brine viscosity, $\gamma_b$ & \SI{0.001}{Pa.s}\\
        CO2 viscosity, $\gamma_c$ & \SI{3e-5}{Pa.s}\\
        Gas constant, $\bar{R}$ & \num{8.32} \si{{\cubic\metre.\pascal}\per {\kelvin.\mole}} \\
        CO$_2$ constant, $a$ & \num{.364} \si{\pascal.\metre^6\per\mole^2} \\
        CO$_2$ constant, $b$ & \num{42.67e-6} \si{\metre^3\per\mole}\\
        Critical temperature, $T_c$  & \SI{303.4}{\kelvin}\\
        Gravitational acceleration, $g$ & \SI{10}{m/s^2}\\
        \hline
    \end{tabular}
    \caption{Properties of the solid phase, brine, and CO$_2$.}\label{tab:properties}
\end{table*}

\begin{figure}[htb!]
    	\includegraphics[width=0.6\textwidth]{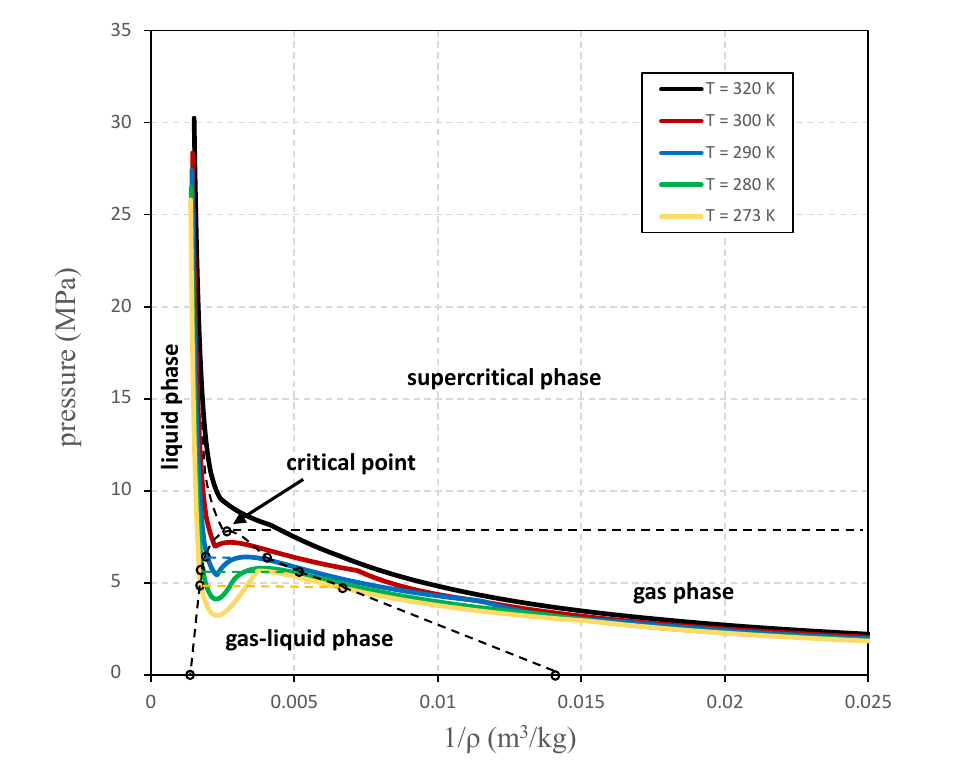}
        \caption{CO$_2$ pressure v. the inverse of CO$_2$ density at different temperatures at $t=$ \SI{10000}{s}. Dashed lines represent the phase boundaries. }
    	\label{fig:fig3}
\end{figure}

Figure \ref{fig:fig3} illustrates the variation of CO$_2$ pressure versus the inverse of CO$_2$ density $\rho_c^{-1}$ at different temperatures, at various points along the horizontal dashed line shown in Figure \ref{fig:fig2}; as expected, it resembles closely the vdW phase diagram.
In this calculation, for simplicity, we do not consider heat transfer between CO$_2$ and the porous medium following ~\cite{goerke2011numerical}, and the temperature consequently refers to the temperature of CO$_2$.
We consider temperatures below and above the CO$_2$ critical temperature ($T_c =$ \SI{303.4}{K}). 
Our results show that close to the injection well and for temperatures below the critical temperature ($T<T_c$), CO$_2$ experiences a phase transition from liquid to gas phase. 
The dashed lines in Figure \ref{fig:fig3} represent the saturated liquid line, critical point, and saturated gas line. 
We find that for lower temperatures, CO$_2$ experiences a sharper change in pressure.

\begin{figure}[htb!]
    	\includegraphics[width=0.6\textwidth]{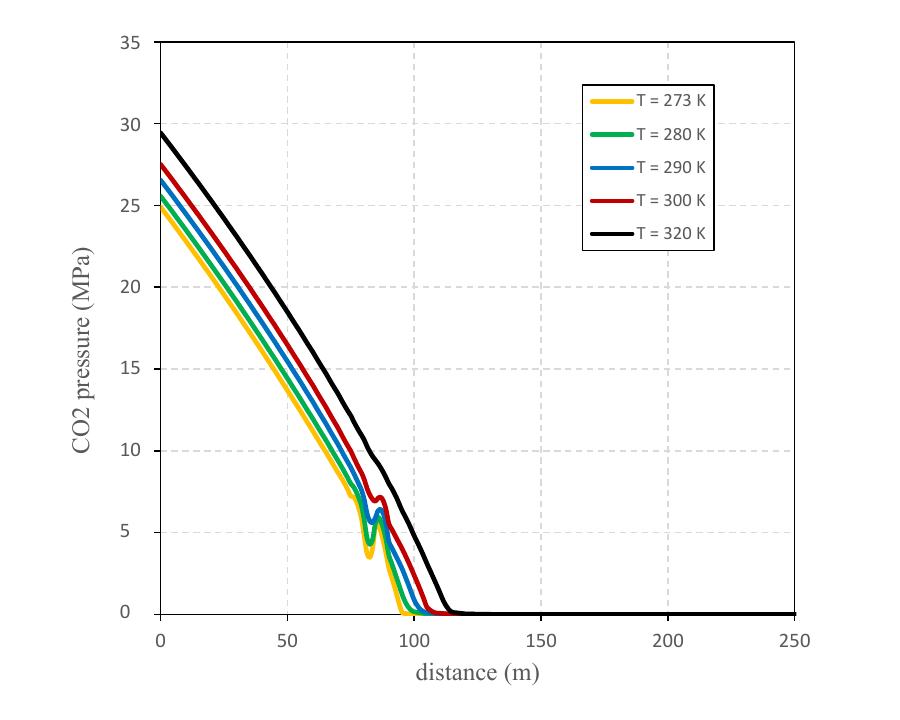}
        \caption{CO$_2$ pressure ($p_c$) v. distance at different temperatures at $t = $ \SI{10000}{s}, and at depth \SI{600}{m}.}
    	\label{fig:fig4}
\end{figure}

Figure \ref{fig:fig4} shows the CO$_2$ pressure versus distance from the injection well for different temperatures at a depth of \SI{600}{m} (the red dashed line shown in Figure \ref{fig:fig2}) and at \SI{10000}{s} after injection initiation. 
Our findings clearly show the phase transition of CO$_2$ close to the injection point for temperature values below the CO$_2$ critical temperature. 
The distance from the well is not sensitive to the injection temperature over the range studied.

\begin{figure}[htb!]
    	\includegraphics[width=0.6\textwidth]{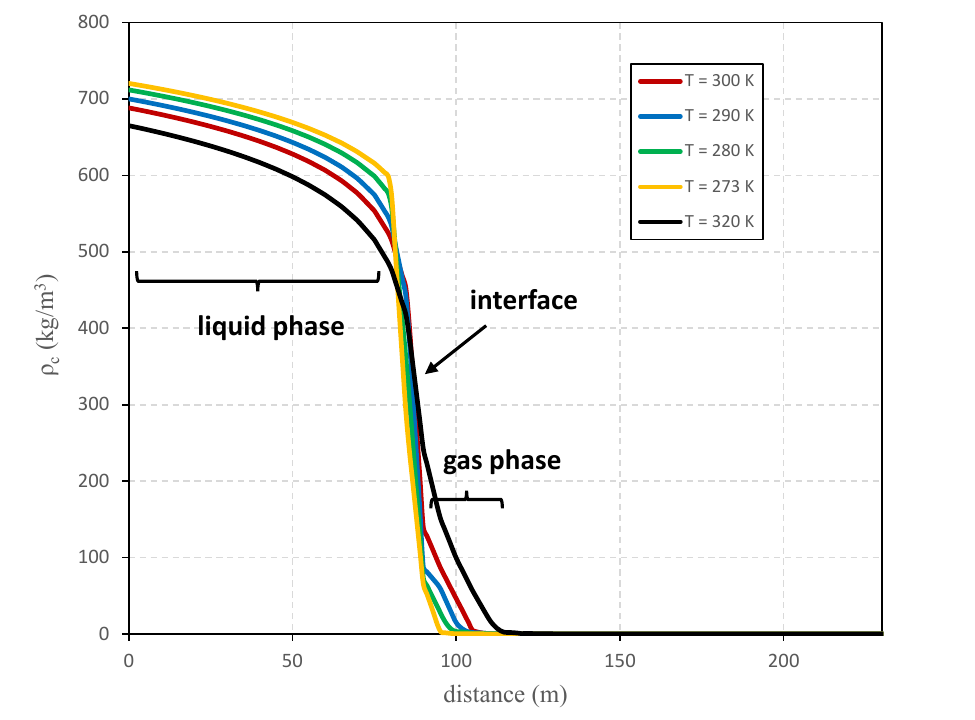}
        \caption{CO$_2$ density $\rho_c$ v. distance for different temperatures at $t = $ \SI{10000}{s}, and at depth \SI{600}{m}.}
    	\label{fig:fig5}
\end{figure}

Figure \ref{fig:fig5} shows the CO$_2$ density $\rho_c$ versus distance from the injection well for different temperatures at \SI{10000}{s} after the start of injection. The abrupt change in the CO$_2$ density highlights clearly the interface between the gas and liquid phases; thus, at  \SI{10000}{s} after injection the interface has migrated to approximately \SI{85}{m} away from the injection well. We find that for lower temperatures, CO$_2$ experiences a larger jump in density. 
Based on this plot, we can precisely locate the phase transition interface at different temperatures.   

\begin{figure}[htb!]
    	\includegraphics[width=0.6\textwidth]{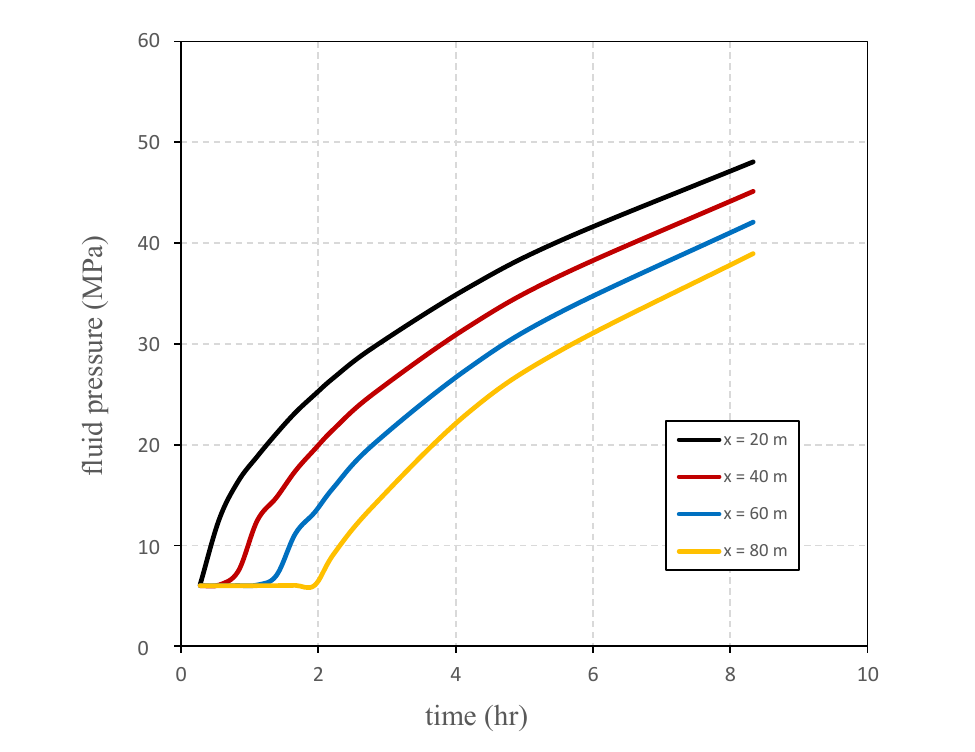}
        \caption{Total fluid pressure v. time at different distances from the injection well at depth \SI{600}{m}.}
    	\label{fig:fig6}
\end{figure}

Figure \ref{fig:fig6} shows the total fluid pressure versus time at distances from the injection well for the monitoring points shown in Figure \ref{fig:fig2}. 
The pressure increases with time throughout the domain near the injection well once the plume migrates to a monitoring location. 
As expected, the points closer to the injection well have higher fluid pressures. 

\begin{figure}[htb!]
    \includegraphics[width=1.0\textwidth]{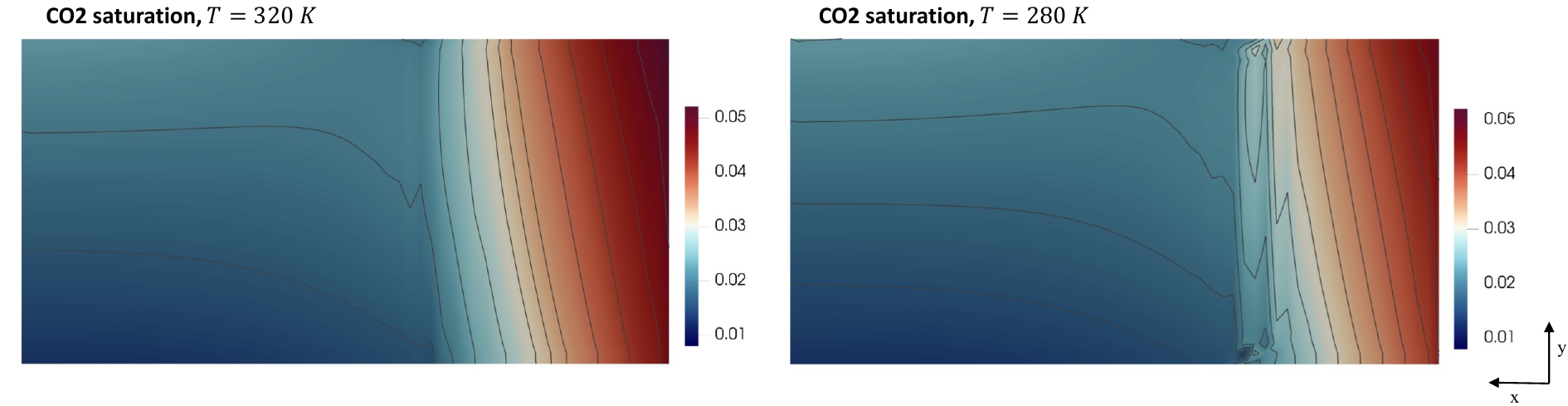}
    \caption{CO$_2$ saturation contours at $t = $ \SI{10000}{s} for temperatures above (\SI{320}{K}, left) and below (\SI{280}{K}, right) the critical temperature.}
    \label{fig:fig7}
\end{figure}

Figure \ref{fig:fig7} shows the CO$_2$ saturation -- defined as $S_c = \phi_c/(1-\phi_s)$ -- contours for two temperatures, one below ($T = $ \SI{280}{K}) and one above ($T = $ \SI{320}{K}) the critical temperature. 
This Figure compares the CO2 migration in the reservoir under super-critical ($T = $ \SI{320}{K}) and sub-critical conditions ($T = $ \SI{280}{K}). 
In this simulation, CO$_2$ in injected to a brine saturated reservoir with initial pressure less than the critical pressure of CO$_2$. In the supercritical case (Fig. \ref{fig:fig7}, left), CO$_2$ undergoes phase transition from supercritical to gas phase, as the migration in the reservoir. In the sub-critical condition (Fig. \ref{fig:fig7}, right) CO$_2$ undergoes a phase transition from liquid to gas phase.
We find that the phase transformation CO$_2$ in the subcritical phase causes to slower propagation in the reservoir.

\subsection{Upward Mobility of CO$_2$ in a Reservoir}
\label{sec:upward}

In this section, we investigate the upward mobility of CO$_2$ in a reservoir by injecting CO$_2$ in the corner of the domain. 
In a geological reservoir, CO$_2$ leakage through faults and movement upward to shallower depths decreases the pressure and temperature of the fluid. 
Therefore, CO$_2$ experiences sub-critical conditions which can result in forming a mixture of gas-liquid phases. 
We consider temperatures both above and below the critical temperature of CO$_2$ to compare the upward mobility of CO$_2$ in the super and sub-critical conditions. 
We find that if CO$_2$ experiences a phase transition, the upward mobility of CO$_2$ decreases.

\begin{figure}[htb!]
    	\includegraphics[width=0.5\textwidth]{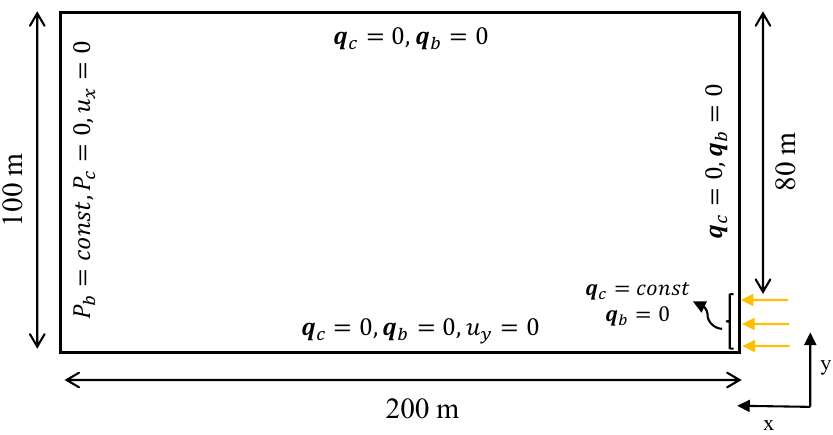}
        \caption{Geometry and boundary condition of the reservoir.}
    	\label{fig:fig8}
\end{figure}

For the simulations, the assumed geometry and boundary conditions are shown in Figure \ref{fig:fig8}. 
We use a constant CO$_2$ injection rate of $q_c = 0.0088$ \SI{}{kg/m^2 s} at the right-bottom corner of the reservoir. 
We consider a $\SI{200}{m} \times \SI{100}{m}$ reservoir, which is initially saturated with brine at an initial pressure of \SI{6}{MPa}.  
The properties of the solid skeleton, brine, and CO$_2$ are from Table \ref{tab:properties}.

\begin{figure}[htb!]
    	\includegraphics[width=1.0\textwidth]{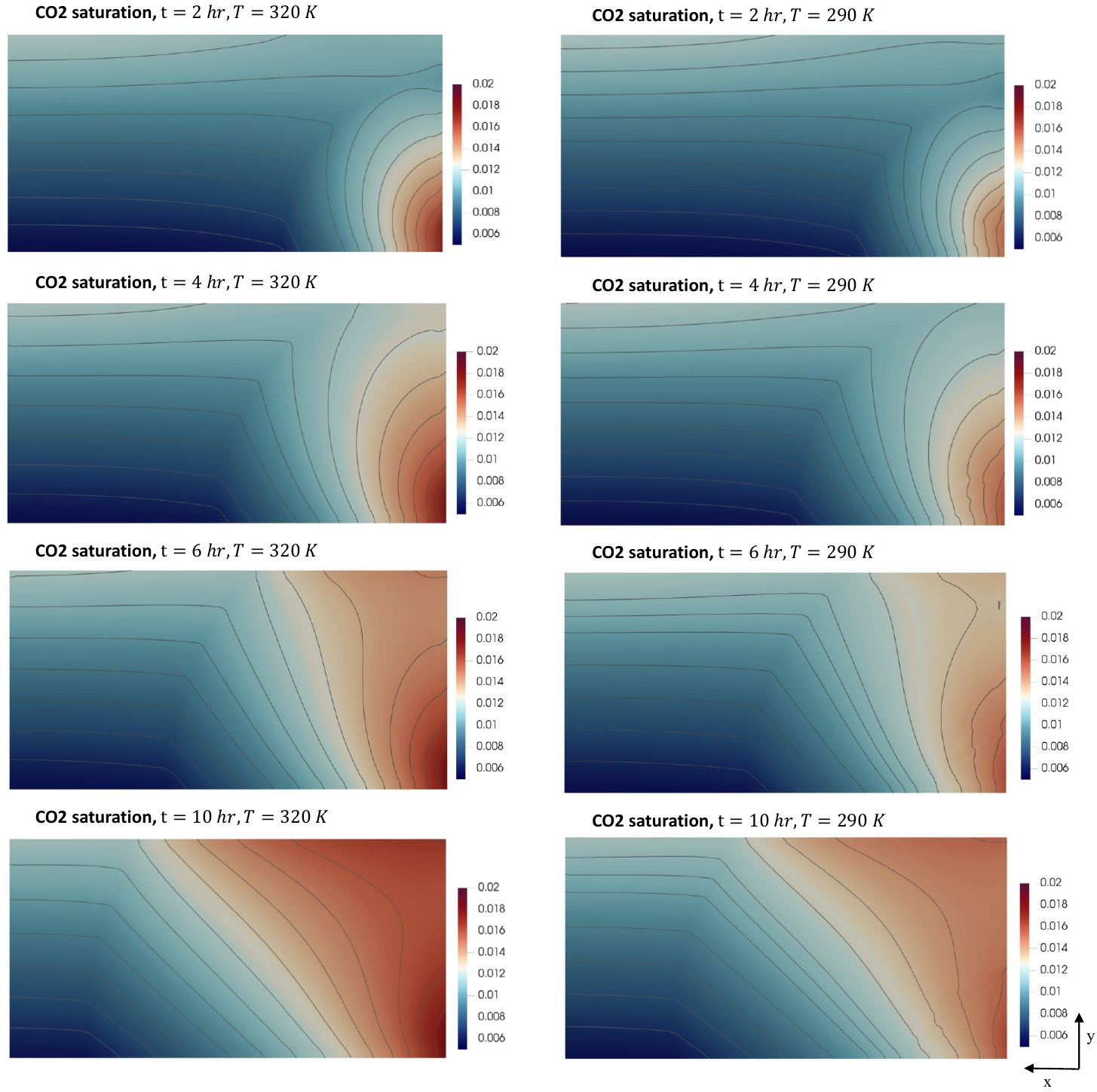}
        \caption{CO$_2$ saturation profiles during injection at super-critical $T = \SI{320}{K}$ (left) and sub-critical $T = \SI{290}{K}$ (right) conditions.}
    	\label{fig:fig11}
\end{figure}

Figure \ref{fig:fig11} compares the simulations considering $T=$\SI{320}{K} (without phase transition), and $T=$\SI{290}{K} (with phase transition) to investigate the effect of CO$_2$ phase change on the upward mobility and migration of CO$_2$ in the reservoir. 
The saturation profiles in Figure \ref{fig:fig11} show that as the injection continues, CO$_2$ migrates to the upper parts of the reservoir due to gravity and the difference in the densities of CO$_2$ and resident fluid (brine). 
At $T=$\SI{320}{K}, when the CO$_2$ is in the super-critical phase, CO$_2$ propagates faster in the domain. 
In the sub-critical condition at $T=$\SI{290}{K}, due to the phase transition of CO$_2$ that forms a gas-liquid mixture and a significant increase of density, we see a slower migration rate and lower value of saturation.

In these simulations, we assumed that the viscosity of CO$_2$ is constant. However, in real soils, the viscosity typically varies, with the liquid phase having a higher viscosity, so it would act to further decrease the mobility of gas-liquid CO$_2$ in comparison to the super-critical phase. 
However, if CO$_2$ undergoes a complete transition to the gas phase under sub-critical conditions, the migration mobility of CO$_2$ will increase in the geological reservoir in comparison to the super-critical phase ~\cite{van2009fluid}. 
We note that, in general, the CO$_2$ saturation depends on the elastic properties of the solid skeleton, permeability, injection rate, and temperature. 

\begin{figure}[htb!]
    	\includegraphics[width=0.6\textwidth]{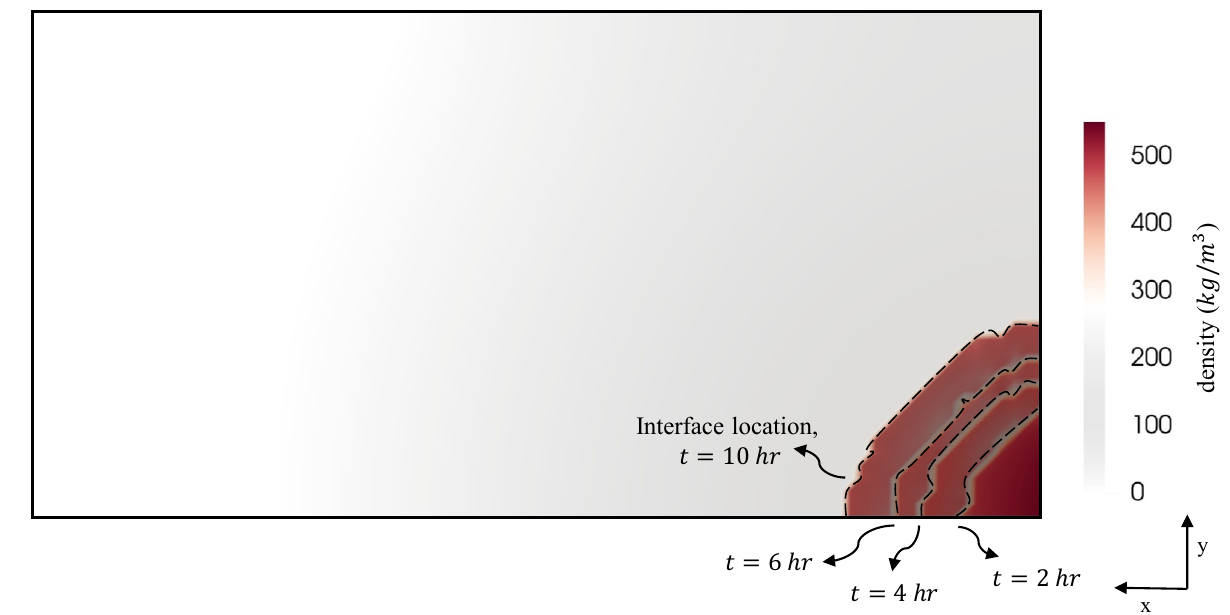}
        \caption{CO$_2$ density profiles and the migrating interface between the CO$_2$ liquid and gas phases during the first $10$ hours of injection, at $T = \SI{290}{K}$.}
    	\label{fig:fig10}
\end{figure}

Figure \ref{fig:fig10} shows the CO$_2$ density profile considering ($T =$\SI{290}{K}), below the critical temperature of CO$_2$. 
We find a clear signature of the migrating interface between the CO$_2$ liquid and gas phases during the injection as it propagates both upward and horizontally.

%%%%%%%%%%%%%%%%%%%%%
%%%%%%%%%%%%%%%%%%%%%
%%%%%%%%%%%%%%%%%%%%%

\section{Concluding Remarks}
In this study, we used a variational energy-based poromechanics model ~\cite{karimi2022energetic} to simulate CO$_2$ sequestration in a porous deformable medium.
An important advantage of the proposed variational formulation is that we can use the van der Waals model for CO$_2$ to model the phase change in a consistent thermomechanical formulation.
It consequently enables us to conduct an investigation of the complex behavior of the CO$_2$ gas-liquid mixture, the moving interface between gas and liquid phases, and the sharp change in the CO$_2$ saturation profile due to the phase change, during the injection and migration into geological formations.
We use the model to compute the pressure, density, and CO$_2$ saturation distributions after injecting dense CO$_2$ into an underground saline formation and examine the liquid-gas CO$_2$ phase transition and its effect on pressure and saturation profiles at different temperatures. 

We studied numerically two model problems that correspond to simplified versions of realistic injection processes.
First, we modeled CO$_2$ injection from a well into an underground saline layer and found that CO$_2$ propagation into the layer slows down if phase transformations occur.
Second, we investigated the upward mobility of CO$_2$ in a reservoir and found that if CO$_2$ experiences a phase transition to form a mixture of gas and liquid phases, the upward mobility of CO$_2$ decreases.
Finally, we note certain limitations of our work and outline potential future directions. First, heat exchange between the fluid phases and the solid porous domain is a significant effect, and the model can be extended beyond the isothermal setting. 
Second, while our current assumptions include a constant viscosity for CO$_2$, the model can be extended by considering viscosity changes associated with phase transformations. 
Finally, although our focus has been on a two-phase flow, the energetic formulation allows us to study unsaturated systems by introducing an additional gas phase.

%%%%%%%%%%%%%%%%%%%%%
%%%%%%%%%%%%%%%%%%%%%
%%%%%%%%%%%%%%%%%%%%%

%%%%%%%%%%%%%%%%%%%%%
%%%%%%%%%%%%%%%%%%%%%
%%%%%%%%%%%%%%%%%%%%%
%%%%%%%%%%%%%%%%%%%%%
\paragraph*{Acknowledgments.}

We thank NSF for XSEDE resources provided by Pittsburgh Supercomputing Center, and John Clayton and Ruby Fu for helpful discussions.
Mina Karimi acknowledges financial support from the Scott Institute for Energy Innovation and the Dowd Fellowship from the College of Engineering at Carnegie Mellon University.
Kaushik Dayal acknowledges financial support from NSF (DMS 2108784) and ARO (MURI W911NF-24-2-0184), and an appointment to the National Energy Technology Laboratory sponsored by the U.S. Department of Energy. 
Matteo Pozzi acknowledges financial support from NSF (CMMI 1638327).
Noel Walkington acknowledges financial support from NSF (DMS 1729478 and DMS 2012259).

\paragraph*{Data Sharing.}

A version of the code developed for this work is available at \url{doi.org/10.5281/zenodo.10820190 }

\paragraph*{Competing Interest Statement.}

The authors declare no competing interest.

%%%%%%%%%%%%%%%%%%%%%
%%%%%%%%%%%%%%%%%%%%%
%%%%%%%%%%%%%%%%%%%%%
%%%%%%%%%%%%%%%%%%%%%
\appendix

\makeatletter
\renewcommand*{\thesection}{\Alph{section}}
\renewcommand*{\thesubsection}{\thesection.\arabic{subsection}}
\renewcommand*{\p@subsection}{}
\renewcommand*{\thesubsubsection}{\thesubsection.\arabic{subsubsection}}
\renewcommand*{\p@subsubsection}{}
\makeatother

%%%%%%%%%%%%%%%%%%%%%
%%%%%%%%%%%%%%%%%%%%%
%%%%%%%%%%%%%%%%%%%%%
%%%%%%%%%%%%%%%%%%%%%

%%%%%%%%%%%%%%%%%%%%%
%%%%%%%%%%%%%%%%%%%%%
%%%%%%%%%%%%%%%%%%%%%
%%%%%%%%%%%%%%%%%%%%%
\appendix
\section{The Peng-Robinson Model for CO$_2$} \label{sec:peng}

In this section, we demonstrate the ability of the approach to handle different free energy functions.
Specifically, we use the Peng-Robinson free energy \cite{poling2001properties} to simulate CO$_2$.
The Peng-Robinson has a better fit than vdW to the properties of CO$_2$, at the cost of introducing more empirical features. 
\begin{align}\label{eq:pen-robinson}
    W_{0c} (\mathcal{R}_{0c}, {\phi}_c, J) &=c \mathcal{R}_{0c} \bar{R}T\left( 1-\log(c\bar{R}T)\right)- \mathcal{R}_{0c} \bar{R}T\log\left( \frac{J{\phi}_c}{\mathcal{R}_{0c}}-b\right) - \frac{a \alpha \mathcal{R}_{0c}}{2b\sqrt{2}} \log \left( \frac{\sqrt{2} + \left(\frac{b \mathcal{R}_{0c}}{J\phi_c} - 1 \right) }{ \sqrt{2} - \left(\frac{b \mathcal{R}_{0c}}{J\phi_c} - 1 \right) } \right)
\end{align}
where $\bar{R}$ is the ideal gas constant, $c$ is a non-dimensional constant, $a$, $b$ and $\alpha$ are constants that relate to the phase transition and are defined as follows:
\begin{gather} \nonumber
    a = 0.4572 \frac{\bar{R}^2 T^2_c}{P_c}, \quad b = 0.0778 \frac{\bar{R} T_c}{P_c}, \quad \alpha = \left( 1+m \left( 1-\sqrt{\frac{T}{T_c}} \right) \right), \text{ where } m = 0.3746 + 1.5422 w - 0.2699 w^2
\end{gather}
where $T_c$ and $P_c$ are critical temperature and pressure of CO$_2$, respectively, and $w$ is the acentric factor of of the fluid.

\begin{figure}[htb!]
    	\includegraphics[width=1.0\textwidth]{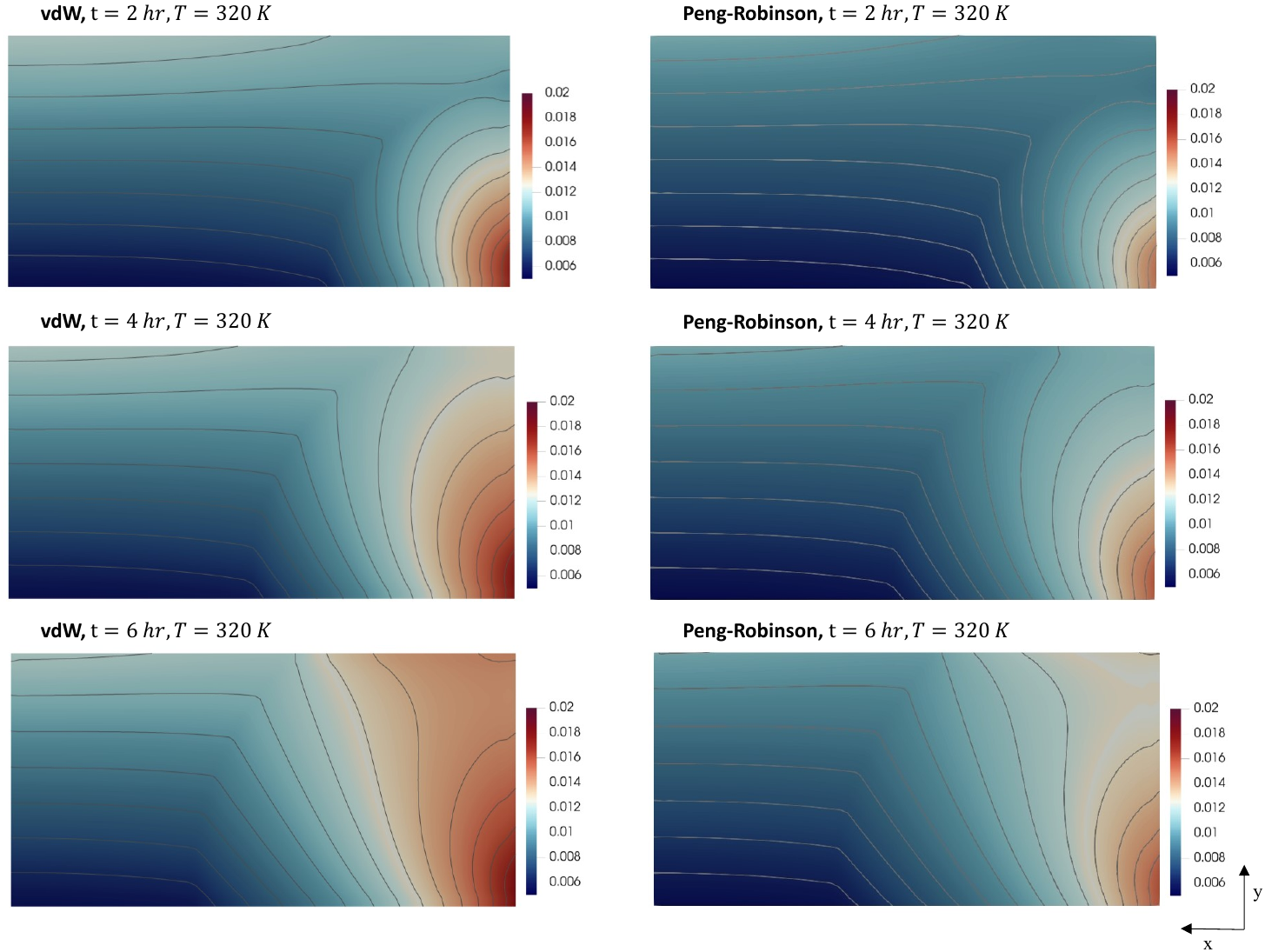}
        \caption{Comparison of CO$_2$ saturation profiles using the vdW model (right) and the Peng-Robinson model (left), at a temperature $T = 320 ~ K$. }
    	\label{fig:peng}
\end{figure}

We substitute the Peng-Robinson free energy \eqref{eq:pen-robinson} in the Lagrangian functional \eqref{eq:Lagrangian}, and obtain the chemical potential of CO$_2$ by taking the variational derivative of Lagrangian with respect to $\mathcal{R}_{0c}$: 
\begin{align}
    \eta_{0c} &= -c\bar{R}T\left( 1-\log (c\bar{R}T) \right) + \bar{R}T\log \left( \frac{1}{\rho_c} -b\right) - \frac{\bar{R}T}{1- b\rho_c} + \frac{a\alpha}{2b\sqrt{2}} \log \left( \frac{\sqrt{2} + \left(b\rho_c - 1 \right) }{ \sqrt{2} - \left(b\rho_c - 1 \right) } \right)
\end{align}
We further obtain the balance of fluid pressure by setting to zero the variational derivative of the Lagrangian with respect to $\tilde{\phi}_c$:
\begin{equation}
    -\frac{RT\rho_c}{1-b\rho_c} + \frac{a\alpha \rho_c^2}{1-b^2\rho_c^2 + 2b\rho_c} + p + p_{0b} = 0
\end{equation}
Figure \ref{fig:peng} compares the saturation profiles resulting from vdW and Peng-Robinson models. In this simulation, the temperature is set to $ = 320 ~ K$, and the material properties and the vdW constants are taken from table \ref{tab:properties}. For the Peng-Robinson model, we assume $w = 0.224$ . The boundary and initial conditions are assumed to be similar to section \ref{sec:upward}. 

%%%%%%%%%%%%%%%%%%%%%
%%%%%%%%%%%%%%%%%%%%%
%%%%%%%%%%%%%%%%%%%%%
%%%%%%%%%%%%%%%%%%%%%
\section{Comparison with Conventional Multiphase Models} \label{sec:comparison}

The movement of CO$_2$ and brine, assumed as two immiscible fluid phases, has been studied using the conventional multiphase method. This method combines an extended form of Darcy's law with mass conservation equations \cite{bear2013dynamics, nordbotten2006similarity}. The multiphase extension of Darcy's law describes the relative velocity of each immiscible phase as follows \cite{court2012applicability, dentz2009abrupt}:
\begin{equation}
    \bfv_i = -\frac{k k_{ri}}{\mu_i} \left( \nabla p_i - \rho_i \bfg \right)
\end{equation}
where $k$ is the true permeability of the medium, and $k_{ri}$ is the relative permeability of each phase. The relative permeability is a function of saturation $S_i$ and varies between zero and one, i.e., $0<k_{ri} (S_i)<1$. The function $k_{ri} (S_i)$ is typically determined empirically based on experimental measurements. For details on different relative permeability relations, we refer to \cite{court2012applicability}. 

The conservation of mass for each fluid phase can be written as
\begin{equation}
    \frac{\partial}{\partial t}\left( \phi_i \rho_i \right) + \divergence (\rho_i \bfv_i) = 0
\end{equation}
where $\phi_i$ is the porosity of each fluid phase and can be defined as $\phi_i = \phi S_i$, where $\phi$ is the porosity of medium and for simplicity can be assumed constant. Under this assumption, the system has six unknowns: $S_c$, $S_b$, $\rho_c$, $\rho_b$, $p_c$, and $p_b$, where subscripts $c$, and $b$ refer to CO$_2$ and brine phases. The system is constrained by the saturation condition $S_c + S_b = 1$. The compressibility conditions for each fluid phase relate the density of each fluid phase to pressure. 
Additionally, the capillary pressure, $p_{\text{capillary}} = p_c - p_b$, provides a relationship between the CO$_2$ and brine pressure. 
$p_{\text{capillary}} (S_i)$ is a function of saturation, which follows empirical relations such as Brooks–Corey \cite{kolditz2012numerical}.

In this work, we neglect the capillary pressure. 
Based on the van der Waals (vdW) free energy, the pressure of CO$_2$ is a function of $\rho_c$. 
We note that our definition of the flux and velocity vectors for each fluid phase (Section \ref{sec:Balance of mass}) leads to a linear relative permeability formulation, given by $k_{ri} = \phi S_i$. 
However, the framework in this paper is general and can be adapted to incorporate capillary pressure effects, nonlinear relative permeability relations, and compressibility of the brine phase, which are topics for future work.


\begin{thebibliography}{10}
	
	\bibitem{NAP25259}
	{National Academies of Sciences Engineering and Medicine}.
	\newblock {\em Negative Emissions Technologies and Reliable Sequestration: A
		Research Agenda}.
	\newblock The National Academies Press, Washington, DC, 2019.
	
	\bibitem{chen2017handbook}
	Wei-Yin Chen, Toshio Suzuki, and Maximilian Lackner.
	\newblock {\em Handbook of climate koli mitigation and adaptation}.
	\newblock Springer International Publishing, 2017.
	
	\bibitem{oldenburg2001process}
	CM~Oldenburg, Karsten Pruess, and Sally~M Benson.
	\newblock Process modeling of {CO2} injection into natural gas reservoirs for
	carbon sequestration and enhanced gas recovery.
	\newblock {\em Energy \& Fuels}, 15(2):293--298, 2001.
	
	\bibitem{ahmad2016injection}
	Nawaz Ahmad, Anders W{\"o}rman, Xavier Sanchez-Vila, Jerker Jarsj{\"o}, Andrea
	Bottacin-Busolin, and Helge Hellevang.
	\newblock Injection of {CO2}-saturated brine in geological reservoir: A way to
	enhanced storage safety.
	\newblock {\em International journal of greenhouse gas control}, 54:129--144,
	2016.
	
	\bibitem{fu2015rock}
	X~Fu, L~Cueto-Felgueroso, D~Bolster, and R~Juanes.
	\newblock Rock dissolution patterns and geochemical shutdown
	of--brine--carbonate reactions during convective mixing in porous media.
	\newblock {\em Journal of Fluid Mechanics}, 764:296--315, 2015.
	
	\bibitem{ilgen2019coupled}
	Anastasia~G Ilgen, Pania Newell, Tomasz Hueckel, D~Nicolas Espinoza, and Manman
	Hu.
	\newblock Coupled chemical-mechanical processes associated with the injection
	of {CO2} into subsurface.
	\newblock In {\em Science of carbon storage in deep saline formations}, pages
	337--359. Elsevier, 2019.
	
	\bibitem{balashov2015reaction}
	Victor~N Balashov, George~D Guthrie, Christina~L Lopano, J~Alexandra Hakala,
	and Susan~L Brantley.
	\newblock Reaction and diffusion at the reservoir/shale interface during {CO2}
	storage: Impact of geochemical kinetics.
	\newblock {\em Applied Geochemistry}, 61:119--131, 2015.
	
	\bibitem{ahmad2016reactive}
	Nawaz Ahmad.
	\newblock {\em REACTIVE TRANSPORT MODELLING OF DISSOLVED {CO2} IN POROUS MEDIA:
		Injection into and leakage from geological reservoirs}.
	\newblock PhD thesis, KTH Royal Institute of Technology, 2016.
	
	\bibitem{wang2023geochemically}
	Jiaan Wang, Wei Xiong, James~B Gardiner, Brandon~C McAdams, Brian~W Stewart,
	R~Burt Thomas, J~Alexandra Hakala, Christina~L Lopano, and Mitchell~J Small.
	\newblock A geochemically informed leak detection (gild) model for {CO2}
	injection sites.
	\newblock {\em Applied Geochemistry}, page 105691, 2023.
	
	\bibitem{karimi2024learning}
	Mina Karimi and Kaushik Bhattacharya.
	\newblock A learning-based multiscale model for reactive flow in porous media.
	\newblock {\em Water Resources Research}, 60(9):e2023WR036303, 2024.
	
	\bibitem{vilarrasa2019induced}
	V{\'\i}ctor Vilarrasa, Jesus Carrera, Sebasti{\`a} Olivella, Jonny Rutqvist,
	and Lyesse Laloui.
	\newblock Induced seismicity in geologic carbon storage.
	\newblock {\em Solid Earth}, 10(3):871--892, 2019.
	
	\bibitem{rutqvist2016fault}
	Jonny Rutqvist, Antonio~P Rinaldi, Frederic Cappa, Pierre Jeanne, Alberto
	Mazzoldi, Luca Urpi, Yves Guglielmi, and Victor Vilarrasa.
	\newblock Fault activation and induced seismicity in geological carbon
	storage--lessons learned from recent modeling studies.
	\newblock {\em Journal of Rock Mechanics and Geotechnical Engineering},
	8(6):789--804, 2016.
	
	\bibitem{chang2019coupled}
	Kyung~Won Chang, Hongkyu Yoon, Mario~J Martinez, and Pania Newell.
	\newblock Coupled hydro-mechanical modeling of injection-induced seismicity in
	the multiphase flow system.
	\newblock In {\em 53rd US Rock Mechanics/Geomechanics Symposium}. OnePetro,
	2019.
	
	\bibitem{newell2020numerical}
	Pania Newell and Mario~J Martinez.
	\newblock Numerical assessment of fault impact on caprock seals during {CO2}
	sequestration.
	\newblock {\em International Journal of Greenhouse Gas Control}, 94:102890,
	2020.
	
	\bibitem{hosseini2012analytical}
	Seyyed~Abolfazl Hosseini, Simon~A Mathias, and Farzam Javadpour.
	\newblock Analytical model for {CO2} injection into brine aquifers-containing
	residual {CH4}.
	\newblock {\em Transport in porous media}, 94(3):795--815, 2012.
	
	\bibitem{huang2015parallel}
	Zhao-Qin Huang, Philip~H Winterfeld, Yi~Xiong, Yu-Shu Wu, and Jun Yao.
	\newblock Parallel simulation of fully-coupled thermal-hydro-mechanical
	processes in {CO2} leakage through fluid-driven fracture zones.
	\newblock {\em International Journal of Greenhouse Gas Control}, 34:39--51,
	2015.
	
	\bibitem{giorgis20072d}
	Thomas Giorgis, Michele Carpita, and Alfredo Battistelli.
	\newblock 2d modeling of salt precipitation during the injection of dry {CO2}
	in a depleted gas reservoir.
	\newblock {\em Energy Conversion and Management}, 48(6):1816--1826, 2007.
	
	\bibitem{preisig2011coupled}
	Matthias Preisig and Jean~H Pr{\'e}vost.
	\newblock Coupled multi-phase thermo-poromechanical effects. case study: {CO2}
	injection at in salah, algeria.
	\newblock {\em International Journal of Greenhouse Gas Control},
	5(4):1055--1064, 2011.
	
	\bibitem{sun2018flow}
	Fengrui Sun, Yuedong Yao, Xiangfang Li, Guozhen Li, Yanan Miao, Song Han, and
	Zhili Chen.
	\newblock Flow simulation of the mixture system of supercritical {CO2} \&
	superheated steam in toe-point injection horizontal wellbores.
	\newblock {\em Journal of Petroleum Science and Engineering}, 163:199--210,
	2018.
	
	\bibitem{negara2011simulation}
	Ardiansyah Negara, MOHAMED~FATHY El-Amin, and SHUYU Sun.
	\newblock Simulation of {CO2} plume in porous media: consideration of
	capillarity and buoyancy effects.
	\newblock {\em International Journal of Numerical Analysis and Modeling, Series
		B}, 2(4):315--337, 2011.
	
	\bibitem{rabinovich2015upscaling}
	Avinoam Rabinovich, Kasama Itthisawatpan, and Louis~J Durlofsky.
	\newblock Upscaling of {CO2} injection into brine with capillary heterogeneity
	effects.
	\newblock {\em Journal of Petroleum Science and Engineering}, 134:60--75, 2015.
	
	\bibitem{song2014analytical}
	Hongqing Song, Gang Huang, Tianxin Li, Yu~Zhang, and Yu~Lou.
	\newblock Analytical model of {CO2} storage efficiency in saline aquifer with
	vertical heterogeneity.
	\newblock {\em Journal of Natural Gas Science and Engineering}, 18:77--89,
	2014.
	
	\bibitem{nordbotten2010analysis}
	Jan~M Nordbotten and Michael~A Celia.
	\newblock Analysis of plume extent using analytical solutions for {{CO2}}
	storage.
	\newblock \url{https://dataspace.princeton.edu/handle/88435/dsp01z603qx41q},
	2010.
	
	\bibitem{okwen2011temporal}
	Roland~T Okwen, Mark~T Stewart, and Jeffrey~A Cunningham.
	\newblock Temporal variations in near-wellbore pressures during {CO2} injection
	in saline aquifers.
	\newblock {\em International Journal of Greenhouse Gas Control},
	5(5):1140--1148, 2011.
	
	\bibitem{vilarrasa2016two}
	V{\'\i}ctor Vilarrasa, Jes{\'u}s Carrera, and Sebasti{\`a} Olivella.
	\newblock Two-phase flow effects on the {CO2} injection pressure evolution and
	implications for the caprock geomechanical stability.
	\newblock {\em E3S Web Conf.}, 9:04007, 2016.
	
	\bibitem{piao2018dynamic}
	Jize Piao, Weon~Shik Han, Sungwook Choung, and Kue-Young Kim.
	\newblock Dynamic behavior of {CO2} in a wellbore and storage formation:
	wellbore-coupled and salt-precipitation processes during geologic {CO2}
	sequestration.
	\newblock {\em Geofluids}, 2018, 2018.
	
	\bibitem{kim2018co2}
	Kiseok Kim, Victor Vilarrasa, and Roman~Y Makhnenko.
	\newblock {CO2} injection effect on geomechanical and flow properties of
	calcite-rich reservoirs.
	\newblock {\em Fluids}, 3(3):66, 2018.
	
	\bibitem{kim2021numerical}
	Taehyun Kim, Chan-Hee Park, Norihiro Watanabe, Eui-Seob Park, Jung-Wook Park,
	Yong-Bok Jung, and Olaf Kolditz.
	\newblock Numerical modeling of two-phase flow in deformable porous media:
	application to {CO2} injection analysis in the otway basin, australia.
	\newblock {\em Environmental Earth Sciences}, 80(3):1--15, 2021.
	
	\bibitem{singh2011non}
	A-K Singh, N~B{\"o}ttcher, W~Wang, C-H Park, U-J G{\"o}rke, and O~Kolditz.
	\newblock Non-isothermal effects on two-phase flow in porous medium: {CO2}
	disposal into a saline aquifer.
	\newblock {\em Energy Procedia}, 4:3889--3895, 2011.
	
	\bibitem{ateshian2022continuum}
	Gerard~A Ateshian and Jay~J Shim.
	\newblock Continuum thermodynamics of the phase transformation of thermoelastic
	fluids.
	\newblock {\em arXiv preprint arXiv:2207.14158}, 2022.
	
	\bibitem{hu2022direct}
	Tianyi Hu, Hao Wang, and Hector Gomez.
	\newblock Direct van der waals simulation (dvs) of phase-transforming fluids.
	\newblock {\em arXiv preprint arXiv:2212.01983}, 2022.
	
	\bibitem{karimi2022energetic}
	Mina Karimi, Mehrdad Massoudi, Noel Walkington, Matteo Pozzi, and Kaushik
	Dayal.
	\newblock Energetic formulation of large-deformation poroelasticity.
	\newblock {\em International Journal for Numerical and Analytical Methods in
		Geomechanics}, 2022.
	
	\bibitem{kolditz2012numerical}
	Olaf Kolditz, Sebastian Bauer, Norbert B{\"o}ttcher, Derek Elsworth, U-J
	G{\"o}rke, C-I McDermott, C-H Park, Ashok~K Singh, J~Taron, and Wenqing Wang.
	\newblock Numerical simulation of two-phase flow in deformable porous media:
	Application to carbon dioxide storage in the subsurface.
	\newblock {\em Mathematics and Computers in Simulation}, 82(10):1919--1935,
	2012.
	
	\bibitem{goerke2011numerical}
	U-J Goerke, C-H Park, W~Wang, AK~Singh, and O~Kolditz.
	\newblock Numerical simulation of multiphase hydromechanical processes induced
	by {CO2} injection into deep saline aquifers.
	\newblock {\em Oil \& Gas Science and Technology--Revue d’IFP Energies
		nouvelles}, 66(1):105--118, 2011.
	
	\bibitem{pruess2004numerical}
	Karsten Pruess.
	\newblock Numerical simulation of {CO2} leakage from a geologic disposal
	reservoir, including transitions from super-to subcritical conditions, and
	boiling of liquid {CO2}.
	\newblock {\em Spe Journal}, 9(02):237--248, 2004.
	
	\bibitem{pruess2011integrated}
	Karsten Pruess.
	\newblock Integrated modeling of {CO2} storage and leakage scenarios including
	transitions between super-and subcritical conditions, and phase change
	between liquid and gaseous {CO2}.
	\newblock {\em Greenhouse Gases: Science and Technology}, 1(3):237--247, 2011.
	
	\bibitem{pruess2011eco2m}
	Karsten Pruess.
	\newblock E{CO2}m: a tough2 fluid property module for mixtures of water,
	{NaCl}, and {CO2}, including super-and sub-critical conditions, and phase
	change between liquid and gaseous {CO2}.
	\newblock Technical report, Lawrence Berkeley National Lab.(LBNL), Berkeley, CA
	(United States), 2011.
	
	\bibitem{lu2014transient}
	Meng Lu and Luke~Daulton Connell.
	\newblock Transient, thermal wellbore flow of multispecies carbon dioxide
	mixtures with phase transition during geological storage.
	\newblock {\em International Journal of Multiphase Flow}, 63:82--92, 2014.
	
	\bibitem{wan2021modeling}
	Nian-Hui Wan, Li-Song Wang, Lin-Tong Hou, Qi-Lin Wu, and Jing-Yu Xu.
	\newblock Modeling transient flow in {CO2} injection wells by considering the
	phase change.
	\newblock {\em Processes}, 9(12):2164, 2021.
	
	\bibitem{sasaki2011heat}
	Kyuro Sasaki and Yuichi Sugai.
	\newblock {\em Heat transfer and phase change in deep {CO2} injector for {CO2}
		geological storage}.
	\newblock INTECH Open Access Publisher, 2011.
	
	\bibitem{van2009fluid}
	LGH~Bert van~der Meer, Cor Hofstee, and Bogdan Orlic.
	\newblock The fluid flow consequences of {CO2} migration from 1000 to 600
	metres upon passing the critical conditions of {CO2}.
	\newblock {\em Energy Procedia}, 1(1):3213--3220, 2009.
	
	\bibitem{vilarrasa2017thermal}
	Victor Vilarrasa and Jonny Rutqvist.
	\newblock Thermal effects on geologic carbon storage.
	\newblock {\em Earth-science reviews}, 165:245--256, 2017.
	
	\bibitem{naghibzadeh2025accretion}
	S~Kiana Naghibzadeh, Anthony Rollett, Noel Walkington, and Kaushik Dayal.
	\newblock Accretion and ablation in deformable solids using an eulerian
	formulation: A finite deformation numerical method.
	\newblock {\em Journal of the Mechanics and Physics of Solids}, 200:106076,
	2025.
	
	\bibitem{gajo2010general}
	Alessandro Gajo.
	\newblock A general approach to isothermal hyperelastic modelling of saturated
	porous media at finite strains with compressible solid constituents.
	\newblock {\em Proceedings of the Royal Society A: Mathematical, Physical and
		Engineering Sciences}, 466(2122):3061--3087, 2010.
	
	\bibitem{li2004dynamics}
	Chao Li, Ronaldo~I Borja, and Richard~A Regueiro.
	\newblock Dynamics of porous media at finite strain.
	\newblock {\em Computer methods in applied mechanics and engineering},
	193(36-38):3837--3870, 2004.
	
	\bibitem{hong2008theory}
	Wei Hong, Xuanhe Zhao, Jinxiong Zhou, and Zhigang Suo.
	\newblock A theory of coupled diffusion and large deformation in polymeric
	gels.
	\newblock {\em Journal of the Mechanics and Physics of Solids},
	56(5):1779--1793, 2008.
	
	\bibitem{fu2016thermodynamic}
	Xiaojing Fu, Luis Cueto-Felgueroso, and Ruben Juanes.
	\newblock Thermodynamic coarsening arrested by viscous fingering in partially
	miscible binary mixtures.
	\newblock {\em Physical Review E}, 94(3):033111, 2016.
	
	\bibitem{abeyaratne2006evolution}
	Rohan Abeyaratne and James~K Knowles.
	\newblock {\em Evolution of phase transitions: a continuum theory}.
	\newblock Cambridge University Press, 2006.
	
	\bibitem{coussy2004poromechanics}
	Olivier Coussy.
	\newblock {\em Poromechanics}.
	\newblock John Wiley \& Sons, 2004.
	
	\bibitem{chua_deformation_2024}
	Janel Chua, Mina Karimi, Patrick Kozlowski, Mehrdad Massoudi, Santosh
	Narasimhachary, Kai Kadau, George Gazonas, and Kaushik Dayal.
	\newblock Deformation {Decomposition} versus {Energy} {Decomposition} for
	{Chemo}-and {Poro}-{Mechanics}.
	\newblock {\em Journal of Applied Mechanics}, 91(1):014501, 2024.
	\newblock Publisher: American Society of Mechanical Engineers.
	
	\bibitem{penrose1990thermodynamically}
	Oliver Penrose and Paul~C Fife.
	\newblock Thermodynamically consistent models of phase-field type for the
	kinetic of phase transitions.
	\newblock {\em Physica D: Nonlinear Phenomena}, 43(1):44--62, 1990.
	
	\bibitem{de2018disclinations}
	Robert~Buarque de~Macedo, Hossein Pourmatin, Timothy Breitzman, and Kaushik
	Dayal.
	\newblock Disclinations without gradients: A nonlocal model for topological
	defects in liquid crystals.
	\newblock {\em Extreme Mechanics Letters}, 23:29--40, 2018.
	
	\bibitem{agrawal2015dynamic}
	Vaibhav Agrawal and Kaushik Dayal.
	\newblock A dynamic phase-field model for structural transformations and
	twinning: Regularized interfaces with transparent prescription of complex
	kinetics and nucleation. part i: Formulation and one-dimensional
	characterization.
	\newblock {\em Journal of the Mechanics and Physics of Solids}, 85:270--290,
	2015.
	
	\bibitem{agrawal2015dynamic-2}
	Vaibhav Agrawal and Kaushik Dayal.
	\newblock A dynamic phase-field model for structural transformations and
	twinning: Regularized interfaces with transparent prescription of complex
	kinetics and nucleation. part ii: Two-dimensional characterization and
	boundary kinetics.
	\newblock {\em Journal of the Mechanics and Physics of Solids}, 85:291--307,
	2015.
	
	\bibitem{fenicsprojectFEniCS}
	{F}{E}ni{C}{S} --- fenicsproject.org.
	\newblock \url{https://fenicsproject.org/}.
	\newblock [Accessed 18-04-2025].
	
	\bibitem{Strang-CSE}
	Gilbert Strang.
	\newblock {\em Computational Science and Engineering}.
	\newblock Wellesley-Cambridge Press, Philadelphia, PA, 2007.
	
	\bibitem{poling2001properties}
	Bruce~E Poling, John~M Prausnitz, John~P O’connell, et~al.
	\newblock The properties of gases and liquids, 2001.
	
	\bibitem{bear2013dynamics}
	Jacob Bear.
	\newblock {\em Dynamics of fluids in porous media}.
	\newblock Courier Corporation, 2013.
	
	\bibitem{nordbotten2006similarity}
	Jan~M Nordbotten and Michael~A Celia.
	\newblock Similarity solutions for fluid injection into confined aquifers.
	\newblock {\em Journal of Fluid Mechanics}, 561:307--327, 2006.
	
	\bibitem{court2012applicability}
	Benjamin Court, Karl~W Bandilla, Michael~A Celia, Adam Janzen, Mark Dobossy,
	and Jan~M Nordbotten.
	\newblock Applicability of vertical-equilibrium and sharp-interface assumptions
	in co2 sequestration modeling.
	\newblock {\em International Journal of Greenhouse Gas Control}, 10:134--147,
	2012.
	
	\bibitem{dentz2009abrupt}
	Marco Dentz and Daniel~M Tartakovsky.
	\newblock Abrupt-interface solution for carbon dioxide injection into porous
	media.
	\newblock {\em Transport in Porous Media}, 79:15--27, 2009.
	
\end{thebibliography}
\end{document}